\documentclass[fleqn,twoside]{article}
\usepackage[headings]{espcrc2}
\usepackage{graphicx}

\readRCS
$Id: espcrc2.tex,v 1.2 2004/02/24 11:22:11 spepping Exp $
\usepackage{graphicx, balance}
\usepackage[figuresright]{rotating}

\newcommand{\AmS}{{\protect\the\textfont2
  A\kern-.1667em\lower.5ex\hbox{M}\kern-.125emS}}

\title{\textbf{Challenges and Issues in Adapting Web Contents on Small Screen Devices}}

\author{Krishna Murthy A \address[DCSE]{Department of Computer Science, University of Mysore, Mysore, 570 006, India \\~Contact: krishnarjun.research@gmail.com \\},
Suresha \address[DCSE]{Department of Computer Science, University of Mysore, Mysore, 570 006, India\\~ Contact: sureshabm@yahoo.co.in\\},
Anil Kumar K M\address{Sri Jayachamarajendra College of Engineering, Mysore, India Contact: anilkmsjce@yahoo.co.in}}
\runtitle{Challenges and Issues in Adapting Web Contents on Small Screen Devices}
\runauthor{Krishna Murthy A, {\it et al.,}.,}

\begin{document}

\begin{abstract}
In general, Web pages are intended for large screen devices using HTML technology. Admittance of such Web pages on Small Screen Device's (SSD's) like mobile phones, palmtops, tablets, PDA $etc.,$ is increasing with the support of the current wireless technologies. However, SSD's have limited screen size, memory capacity and bandwidth, which makes accessing the Website on SSD's extremely difficult. There are many approaches have been proposed in literature to regenerate HTML Web pages suitable for browsing on SSD's. These proposed methods involve segment the Web page based on its semantic structure, followed by noise removal based on block features and to utilize the hierarchy of the content element to regenerate a page suitable for Small Screen Devices. But World Wide Web consortium stated that, HTML does not provide a better description of semantic structure of the web page contents. To overcome this draw backs, Web developers started to develop Web page(s) using new technologies like XML, Flash $etc..$ It makes a way for new research methods. Therefore, we require an approach to reconstruct these Web pages suitable for SSD's.  However, existing approaches in literature do not perform well for Web pages erected using XML and Flash. In this paper, we have emphasized a few issues of the existing approaches on XML, Flash Datasets and propose an approach that performs better on data set comprising of Flash Web pages. 
\end{abstract}

% typeset front matter (including abstract)
\maketitle

\section{INTRODUCTION}
The rapid expansion of internet has made Web a popular place for dissemination of information and also provided avenues for research in various fields related to the web. In last few decades, research on Web is increasing at rapid rate. For example, improving the quality of Web by Analyzing Usability Test, Web Information Extraction, Tracking Product Opinions by analyzing user reviews, Browsing Web on SSD's. In general it is called as ‘Web Mining’.
\vskip 2mm
According to analysis targets [1], Web mining is divided into three different types namely Web Usage Mining, Web Structure Mining and Web Content Mining. Web usage mining is the process of determining the patterns of users on the internet. It describes how a page is accessed, date and time, the page was accessed, IP address of the browser and page references $etc.$ [2][3]. Web Structure Mining is the process of using graph theory to analyze the node and connection structure of the Web site.  Web Structure Mining can be divided into two kinds; extracting patterns from hyperlinks in the Web, a hyperlink is a structural component that connects the Web page to a different location, and mining the document structure, analysis of the tree link structure of page structures to describe HTML or XML tag pages [4]. Web Content Mining aims in extracting useful information or knowledge from Web page content [5].
\vskip 2mm
Currently surfing the Web on SSD's such as Mobile phones, Personal Digital Assistants (PDA) $etc.,$ is becoming very popular. Delivering Web pages to SSD's have become possible with the advanced wireless technology. But, the current Web pages are intended for Large Screen Device's (LSD's) and are not suitable for SSD's. In literature, it is observed that the straight forward solution for browsing on SSD's is to redesign the Web pages using specific markup languages such as WML, XHTML. Some notable sites, which are already done this, include Yahoo, CNN, and Google among others. Nonetheless, the vast majority of sites on
the Web do not have customized Web pages for SSD's because it's time consuming process and not economical[6]. 
\vskip 2mm
Compared to LSD's, SSD's are not ideal platforms for surfing the Web as the wireless bandwidth is quite limited, it's very expensive and screen size varies for different devices such as mobiles, PDA's $etc.$ and have limited memory capabilities. Normally, the content of a single Web page will be larger than contents that can be accommodated in a mobile phone. Therefore, methods to translate LSD's optimized Web pages for SSD's are essential. Several attempts have been made to enable Web pages to be browsable on SSD's. All the methods designed were based on HTML tags as there is a vast availability of HTML pages on Web [5][7]. World Wide Web (WWW) consortium stated that, HTML has lot of drawbacks such as limited defined tags, case in sensitive, semi-structured and designed to display data with limited options[6]. To overcome these difficulties a few technologies were introduced such as XML [8][9], Flash $etc.$. As a result, majority of the Web pages are developed using XML and Flash technologies. These kinds of Web pages are not viable for browsing on SSD's and majority of the information on these kinds of Web pages are not included in the results of Web search engines.
\vskip 2mm		
In our approach, we create a dataset for XML and Flash Web pages and analyze the feasibility of existing Web page segmentation systems. We are working on approaches to translate XML and Flash Web pages for efficient browsing on SSD's. In this paper, we will highlight issues on adaptation of the existing works on SSD's and also propose a method to translate Flash Web pages browsable on SSD's. This paper is structured as follows. In Section 2, we discuss literature work of potential features for detecting Web page contents, noise removal and informative content adaptation. Section 3 presents methodology of our approach. Section 4 discusses experiments and results analysis of existing and proposed approach. Conclusion is provided in Section 5.

\section{RELATED WORKS}
\label {section:RoL}

SSD's have received a substantial attention from both research community and industry and it still remains a challenge for them to enable SSD users to browse Web pages built using advanced technologies such as Flash, Silver light, XML $etc.$.  Majority of the earlier works concentrated on displaying HTML Web pages on SSD, as earlier Web pages available in world wide Web were created using HTML [10].   
\vskip 2mm
In 2003, Vision Based Page Segmentation (VIPS) algorithm [10] was proposed to extract the semantic structure of a Web page. Semantic structure is a hierarchical structure in which each node will correspond to a block and each node will be assigned a value to indicate degree of coherence based on visual perception. It may not work well and in many cases the weights of visual separators are inaccurately measured [11],   as it does not take into account the Document Object Model (DOM) tree information and when the blocks are not visibly different. Gestalt Theory [11]: a psychological theory that can explain human's visual perceptive process. The four basic laws, Proximity, Similarity, Closure and Simplicity are drawn from Gestalt Theory and then implemented in a program to simulate how human understand the layout of Web pages. 
\vskip 2mm
A graph-theoretic approach [12] is introduced based on formulating an appropriate optimization problem on weighted graphs, where the weights capture if two nodes in the DOM tree should be placed together. Christian {\it et al.,} [7] provided an abstract block-level page segmentation model that focuses on the low-level properties of text instead of DOM-structural information. The key observation is that the number of tokens in a text fragment (or more precisely, its token density) is a valuable feature for segmentation decisions. Liu {\it et al.,} [13] proposed a novel Web page segmentation algorithm based on finding the Gomory-Hu tree in a planar graph. The algorithm initially distills vision and structure information from a Web page to construct a weighted undirected graph, whose vertices are the leaf nodes of the DOM tree and the edges represent the visible position relationship between vertices.  It then partitions the graph with the Gomory-Hu tree based clustering algorithm. 
\vskip 2mm
Unlike conventional data or text, Web pages typically contain a large amount of information that is not art of the main contents of the pages, $e.g.,$ banner, ads, navigation bars, copyright notices and so on. Such irrelevant information's also known as Web page noise in Web pages can seriously harm Web mining tasks such as Web page clustering, Web page classification, Web search engine $etc.$.
\vskip 2mm
By considering the above issues, Ruihua Song {\it et al.,} [14] proposed by a system to formulate the block importance estimation as a learning problem. Here VIPS [10] is applied for segmentation and Spatial features, content features of each blocks are extracted to construct a feature vector for the each block and then learning algorithms such as SVM and Neural Network methods are used to train a model to assign importance to each block. Jing Li and C I Ezeife [9] introduced the system called Webpage Cleaner for eliminating noise blocks from Web pages, it first extracts Web blocks using VIPS [10] then relevant Web page blocks are identified by analyzing physical features of the blocks such as the block location, percentage of Web links on the block and level of similarity of block contents to other blocks. An effective approach for boilerplate (noise) detection using shallow text features such average word length, average sentence length, absolute number of words and link density $etc.$, for classifying the individual text elements in a Web page and then compared the approach to complex, state-of-the-art techniques and shown that competitive accuracy achieved, at almost no cost[15]. Thanda Htwe {\it et al.,} [16] introduced the system to detect multiple noise patterns from Web pages. The method is based on the basic idea of Case Based Reasoning (CBR) to find noise pattern in current Web page by matching similar noise pattern kept in Case-Based. Back propagation Neural Network algorithm was employed to classify the stored various noise patterns by matching similar noise data. 
\vskip 2mm
Most of the existing Web pages are designed for desktop PC's, which makes viewing them on SSD's extremely difficult due to limited bandwidth, small screen and limited memory. Yin X and Lee W S [17] proposed a very first system by using a ranking algorithm similar to Google Page Ranking algorithm to rank the content objects within a Web page. This allows the extraction of only important parts of Web pages for delivery to mobile devices. Che Y {\it et al.,} [18] introduced a new page adaptation technique which analyzes Web page structure and splits it into smaller, logically related units that can fit onto a mobile device screen. The authors first analyzed the HTML DOM tree and detected the high-level content blocks and then analyzed the content inside each high level content block to identify explicit separators to determine splitting of the blocks and finally detect implicit separators to help split the blocks further. The overall analysis is to split Web pages into appropriate blocks so that users can browse page blocks on SSD's.
\vskip 2mm
 Hattori {\it et al.,} [19] introduced the method to reconstruct the PC's optimized Web pages for mobile browsing, the approach followed is to segment the Web pages based on its content distance and utilize the hierarchy of the content element to regenerate a page suitable for mobile phone browsing. In 2009, Xin Yang [20] proposed the novel approach to segment Web pages into mobile-fitted blocks guided by four general laws in Enhanced-Gestalt Theory. This method first group's visually and semantically coherent content into hierarchical parts according to similarity, closure and simplicity laws in E-Gestalt theory and then divides them into mobile fitted blocks using proximity law. Finally through proxy automatically re-author HTML documents into mobile-intended structures using segmentation results.
\vskip 2mm
From literature we have observed that HTML syntax is very flexible, as a lot of HTML Web pages do not comply to the specifications put forth by World Wide Web Consortium.  It has a lot of drawbacks such as limited defined tags, case insensitive, semi-structured and designed to display data with limited options, which will cause problem in DOM tree structure [6]. To provide better descriptions of the semantic structure of the Web page content, a few new technologies Flash, XML etc., are introduced. However, a majority of the Web pages are in HTML rather than the other technologies. Therefore the existing works reported in literature have focused on only HTML Web pages [5]. There is no work reported in literature to segment, remove noise and adapt contents on SSD's for Flash and XML related Web page's.

\section{METHODOLOGY}

We began the process of creating data set for XML and Flash Web pages in order to study the viability of the existing approaches. Since there do not exist such data sets, we have proposed a method to crawl URL's from given Web domains and to classify the uniform resource locations (URLs) comprising of XML Web pages using string based search [21]. We have collected around 120 million Web domain URL's from different Web domains like .org, .com, .net, .info, .us and .sk. Bunch of Web domains from each domain has provided as input to our crawler to extract XML URL's. our crawler extracts around 2400 XML URL's. Here we considered this bulk XML URL's as Dataset-1 for experimentations. Similarly we have created a dataset for Flash URL's manually based on its semantic analysis, as it does not contain any identical/unique extension such as .html, .xml, .aspx and so on. Here we have collected nearly 80 various Flash Web domains containing approximately 300 URL Web pages for the exper-imentation (Dataset-2).
\vskip 2mm
In order to study viability of existing approaches on SSD's, we have implemented a few existing approaches that predominantly deal with segmentation for displaying Web contents on SSD's. In this section, we have described the working of two such popular approaches namely VIPS [10] and Boilerpipe systems [15] on the two Datasets. 
\vskip 2mm
VIPS is a Web content structure analyzer or Web page analyzer based on visual representation algorithm. Many Web applications such as information retrieval, information extraction and automatic page adaptation can benefit from this structure [10]. It discusses about an automatic top-down, tag-tree independent approach to detect Web content structure and simulates user understanding of Web layout structure based on visual perception. After setting-up VIPS system, we tested system feasibility on our dataset's, and found that these pages does not get segmented. Figure 1 depicts successful segmentation of normal HTML Web page (ex: http://www.uni-mysore.ac.in) after applying PDoC value (Permitted Degree of Coherence) which controls the partition granularity; here the left part shows the segmented blocks (different visual blocks). The block marked with red rectangle is the selected content structure VB1-1-3(4). The upper right part shows the PDoC value and vision based content structure (hierarchical structure of different blocks – VIPS tree) of Web page, bottom right part gives the features of selected content structure. 

%\begin{figure}[ht]
%\centering
%\includegraphics[height=3cm]{1eps}
%\caption{VIPS page analyzer on HTML Web page.}
%\end{figure}
%\onecolumn
\begin{figure*}[!ht]
%\parbox{5.5cm}{
\centering
\includegraphics[width=12cm, height=6cm]{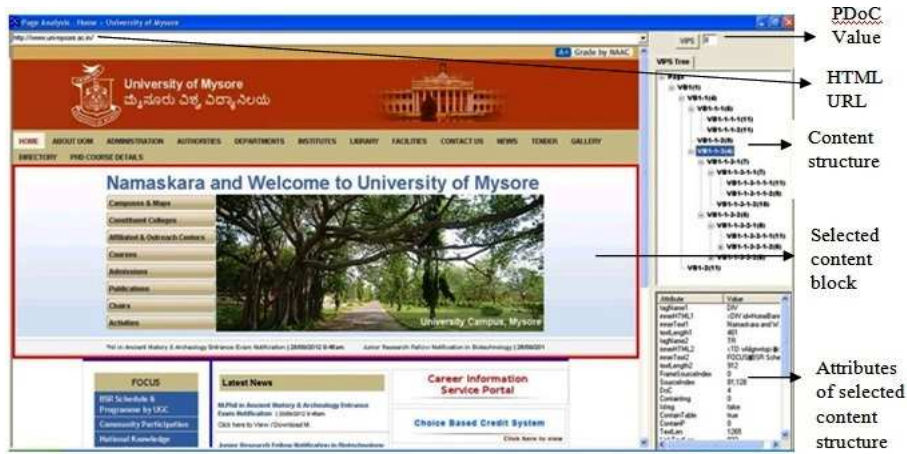}\\
\caption{VIPS on HTML Web page}
\end{figure*}

%\begin{figure*}[!ht]
  % Requires \usepackage{graphicx}
    %\scalebox{0.4}
    %{\includegraphics[angle=0]{CompressedDomainProcessing.jpg}}
 %   \centering
  %  \includegraphics[height=75pt,width=75pt]{sampleDocument.jpg}\\
  %  \caption{Scanned sample document }
  %  \label{Fig:sampleDocument}
 %\end{figure*}

\begin{figure*}[!ht]
%\parbox{5.5cm}{
\centering
\includegraphics[width=12cm, height=6cm]{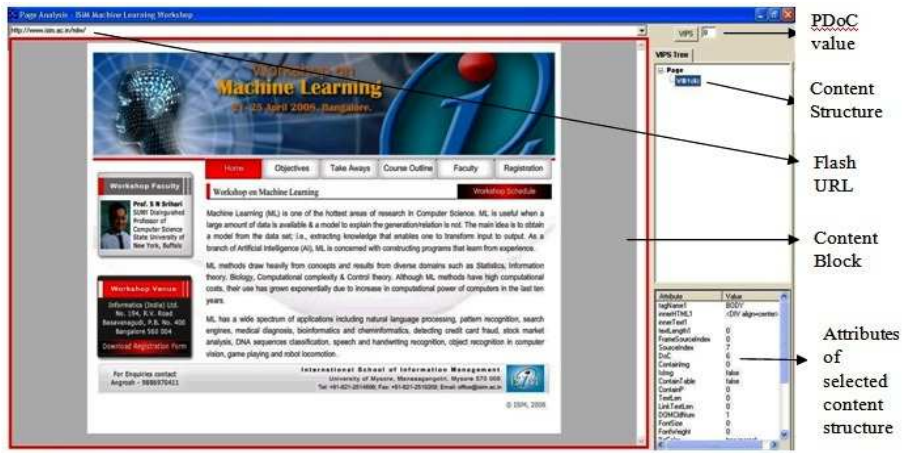}\\
\caption{VIPS Flash Web page.}
\end{figure*}
\vskip 2mm
In different applications, we can control the partition granularity by setting PDoC value. Whereas VIPS system fails to segment Flash Web pages (ex: http://www.isim.ac.in/mlw) after entering PDoC value. Figure 2 depicts failure of segmentation of Flash Web pages after applying PDoC value. It segments entire Web page as a single block, but it fails to segment as individual blocks. The reason behind this, Flash Web pages are semantically different from HTML pages. VIPS is developed based on pre-defined HTML tags.
Figure 12 and Figure 11 depict the feasibility analysis of VIPS systems on various kind of Webpages (HTML, Flash and XML repsectively). 
 
%\begin{figure}[ht]
%\centering
%\includegraphics[height=3.5cm]{2eps}
%\caption{VIPS page analyzer on Flash Web page.}
%\end{figure}

\begin{figure*}[!ht]
%\parbox{5.5cm}{
\centering
\includegraphics[width=12cm, height=6cm]{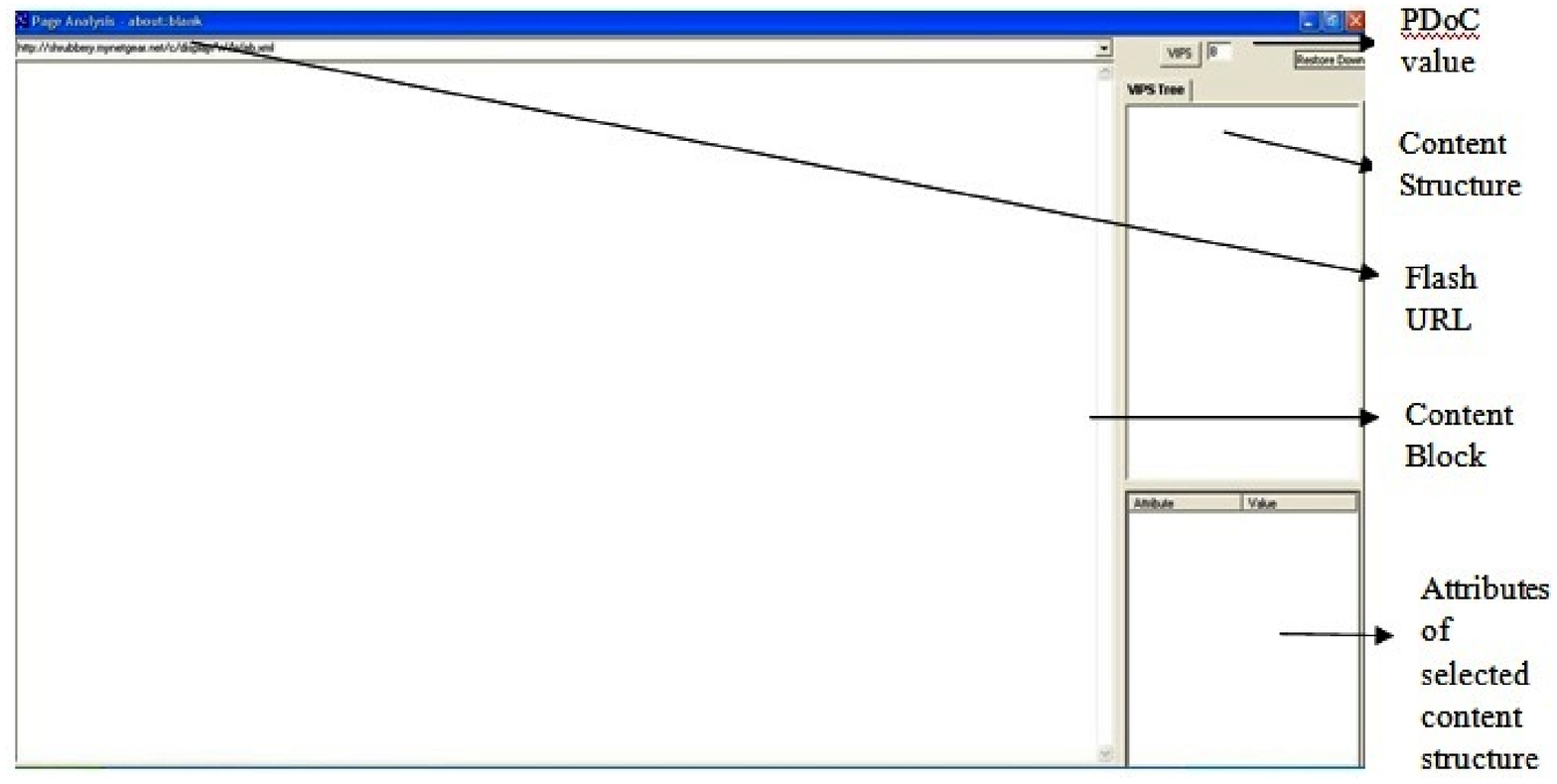}
\caption{VIPS on XML Web page.}
\end{figure*}

\begin{figure*}[!ht]
%\parbox{5.5cm}{
\centering
\includegraphics[width=12cm, height=6cm]{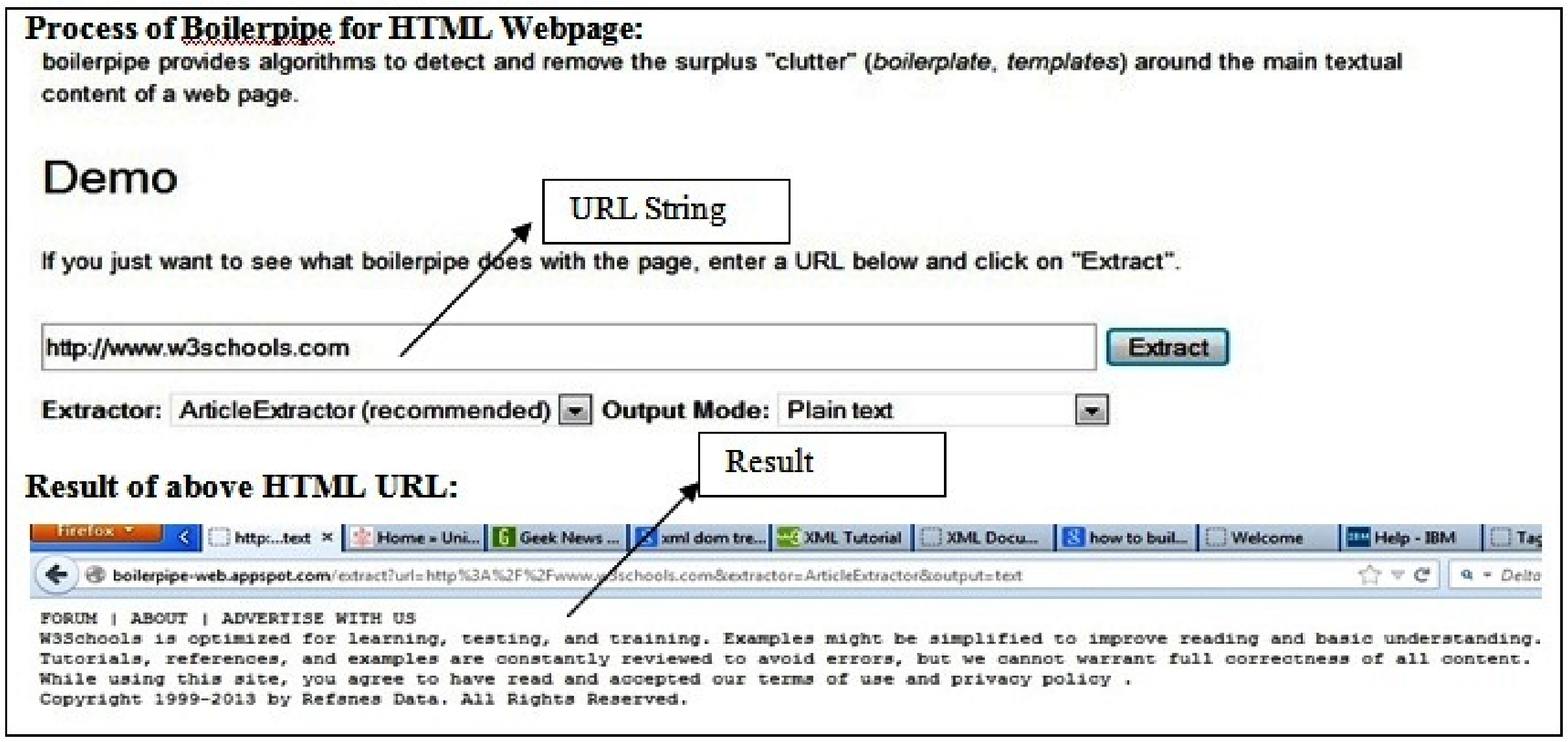}
\caption{Boilerpipe page analysis on HTML Web page.}
\end{figure*}

\vskip 2mm
Similarly, we have tested on XML Dataset.  VIPS system fails to display the Web contents in display region. Figure 3 depicts the pictorial view of XML (ex: http://shrubbery.mynetgear.net/c/disply/W/ Web.xml) Web pages on VIPS system. The reason of failure is that the semantic structure of XML Web pages is different from HTML Web pages. Hence, we have concluded that the existing VIPS model fails to segment the Flash Webpage.

%\begin{figure}[ht]
%\centering
%\includegraphics[height=3.5cm]{3eps}
%\caption{VIPS page analyzer on XML Web page.}
%\end{figure}

\vskip 2mm
Boilerpipe provides algorithms to segment and remove the surplus “clutter” (boilerplate, templates) around the main textual content of a Web page [15].  We have tested the Boilerpipe model on our datasets, for all cross combinations of extractor and output mode. They are employed to segment and retrieve informative contents from Web pages. The model works fine for HTML Web pages, but fails to segment and extract informative contents from Flash and XML Web pages. The model shows the result as blank screen or returns ‘code error’ message. Figure 4 depicts successful segmentation and information extraction of HTML Web page. Figure 5 and Figure 6 depict the failure of segmentation and information extraction of Flash and XML Web pages respectively after applying Extractor and Output mode options of Boilerpipe model.

\begin{figure*}[!ht]
\centering
\includegraphics[width=12cm, height=6cm]{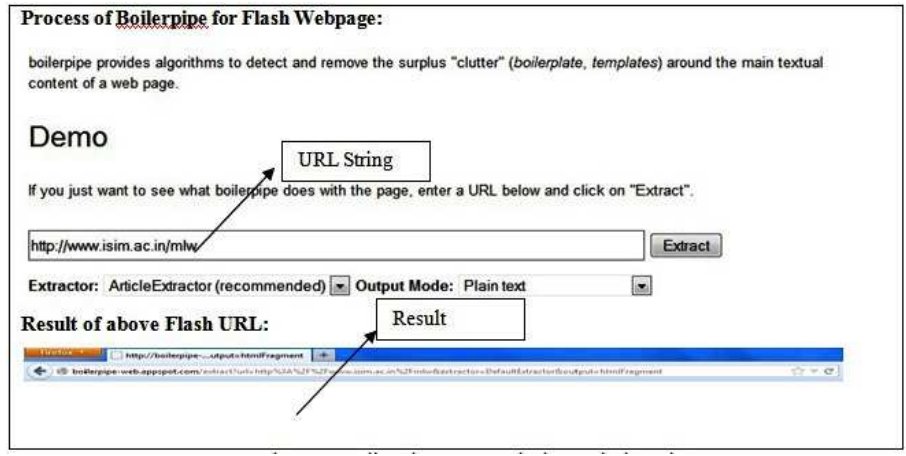}
\caption{Boilerpipe page analysis on Flash Web page.}
\end{figure*}

\begin{figure*}[ht!]
\centering
\includegraphics[width=12cm, height=6cm]{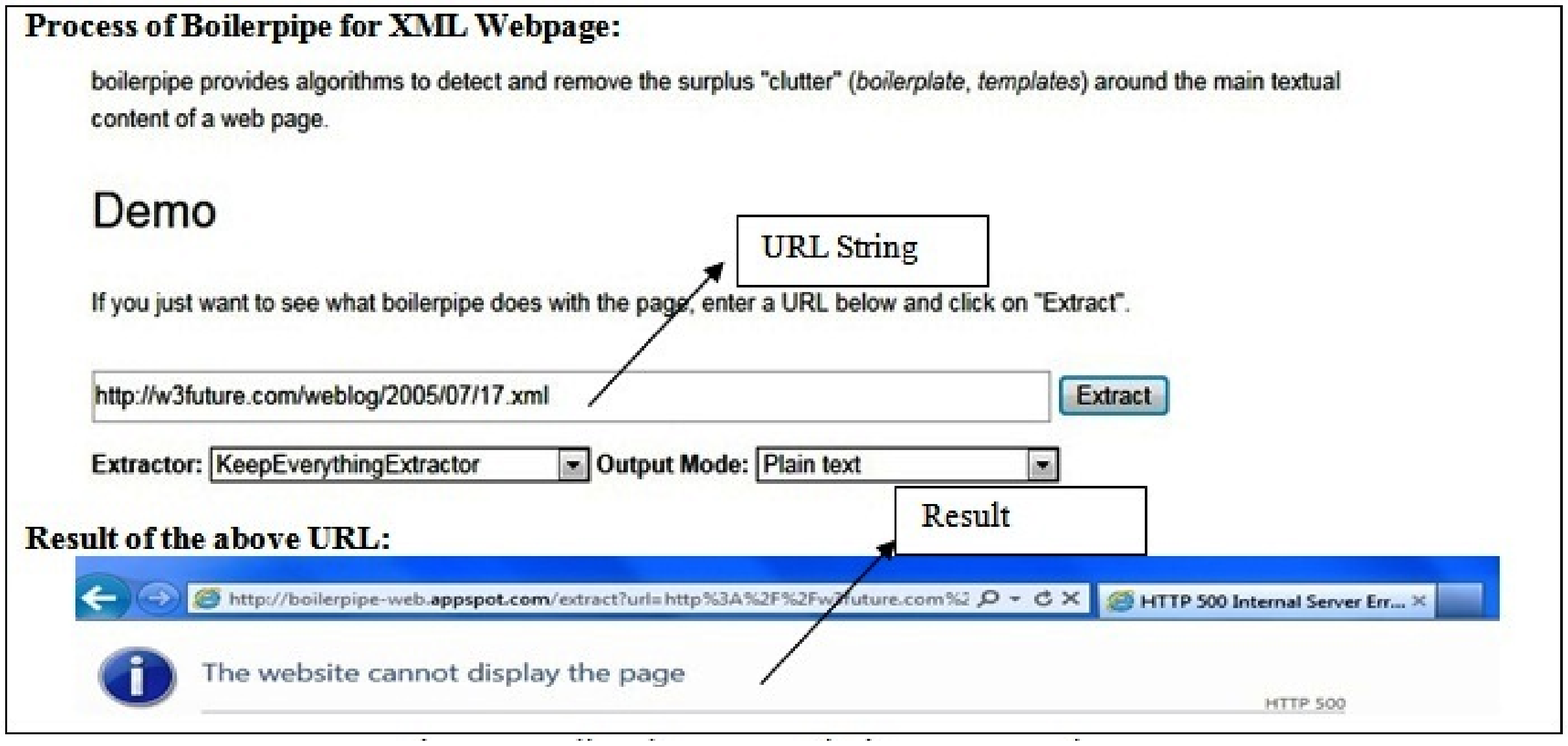}
\caption{Boilerpipe page analysis on XML Web page.}
\end{figure*}

%\begin{figure}[ht]
%\centering
%\includegraphics[height=3.6cm]{5}
%\caption{Boilerpipe page analysis on Flash Web page.}
%\end{figure}

%\begin{figure}[ht]
%\centering
%\includegraphics[height=3.6cm]{6}
%\caption{Boilerpipe page analysis on XML Web page.}
%\end{figure}
\begin{figure*}[ht!]
\centering
\includegraphics[width=10cm, height=6cm]{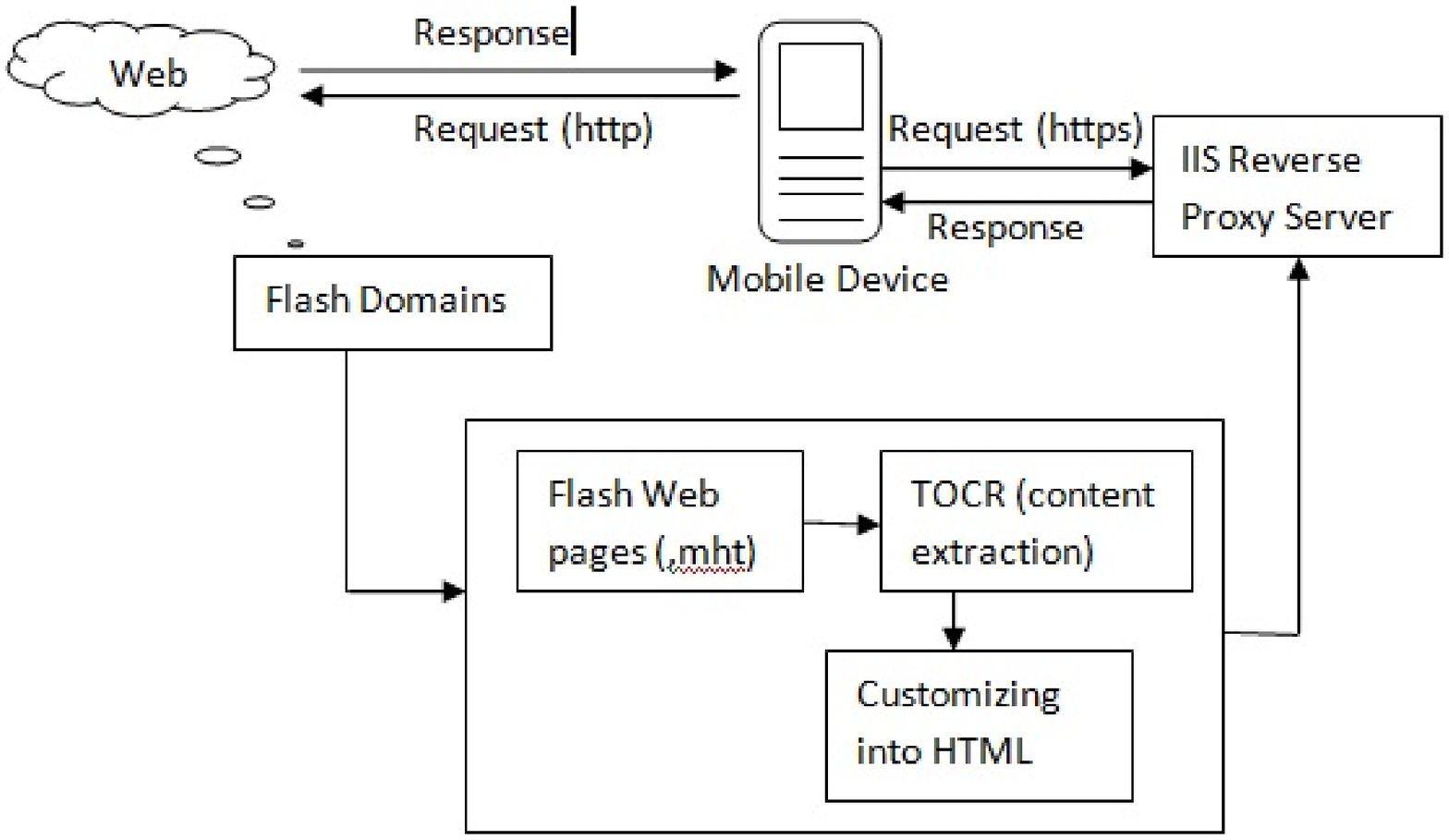}
\caption{Architecture of Proposed System.}
\end{figure*}

\vskip 2mm
From this study, we have learnt that the Boilerpipe model also fails to segment and extract informative contents from Flash and XML Web pages. Therefore from the analysis of the existing systems, we have shown that these systems via segmentation, noise removal and mobile browsing work well on Web pages using pre-defined HTML tags.  Failure of these systems to segment and extract informative contents from Flash and XML Web pages has a few limitations such as inability to display these contents on SSD's, inability to share contents from SSD's to social Web sites, inability to index these Web pages by Web search engines resulting in lack of its use in information retrieval, information extraction etc., citation rate of these Web pages are also reduced due afore mentioned limitations.
\vskip 2mm   
In order to address these limitations, we are working on semantic structure of XML and Flash Web pages to explore the segmentation technique. In this paper, we have proposed an approach for translation of Flash Web pages into HTML format, suitable for Web browsing on SSD's. We have analyzed the browsing performance using two metrics namely content coverage and response time on different mobile phones.

\subsection{Translation of Flash Web pages}
In this approach, after downloading the Flash Web pages .mhtml format, each tab pages are converted into images using snipping tool. The images are saved in bitmap format and text contents of each image extracted by using the Transym Optical Character Recognition (TOCR) tool. Then the corresponding HTML Web pages are constructed similar to original Flash Web pages.
\vskip 2mm
   After the translation of HTML pages, it is hosted using reverse proxy concepts with the help of Internet Information Service (IIS) Web server. A very common reverse proxy scenario is to make available several internal Web applications over the Internet.    An Internet-accessible Web server is used as a reverse-proxy server that receives Web requests and then forwards them to several intranet applications for processing.
\vskip 2mm	

We have setup an Reverse Proxy server using the Application Request Routing and the URL Rewrite Module 2.0. After this concepts, translated Web domains are hosted by assigning an individual port number to respective domain names for easier access. Testing were done on different SSD's based on content coverage and response time. Figure 7 depicts the architecture of the proposed system.

\section{EXPERIMENTS AND RESULTS}

Web pages are hosted after translation into HTML format (traditional Web pages). The results have been analyzed based on accessing the Conventional Flash Web pages (C-Web pages) as well as Traditional Web pages (T-Web pages). Here system is analyzed by comparing the time factors of downloading time on SSD's (based on system specification) and content coverage. We have performed the downloading time analysis on three different specification hand held devices such as Sony Xperia X10, Sony Ericsson WT13i and LG Optimus Net based on Wi-Fi connection. We obtained the better performance in accessing T-Web pages compared to its corresponding C-Web pages.  Figure 13 and Figure 14 provide results of response time and content coverage respectively on our approach on Flash data set.

%\begin{table}[ht]
%\caption{\bf Response Time analysis on various SSD's}
%\label{tab1}
%\end{table}
\begin{figure}[ht!]
%\parbox{5.5cm}{
\centering
\includegraphics[width=7cm, height=5cm]{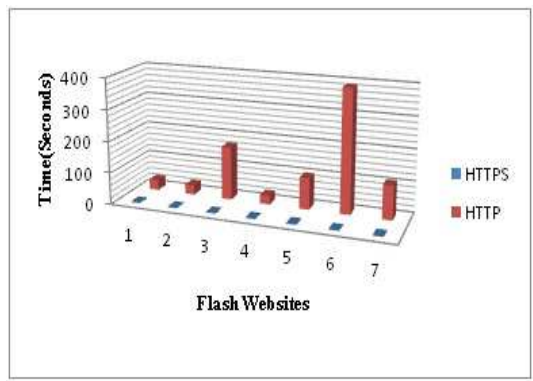}
\caption{Performance Analysis on SonyXPeria10.}
\end{figure}

\begin{figure}[ht!]
\centering
\includegraphics[width=7cm, height=5cm]{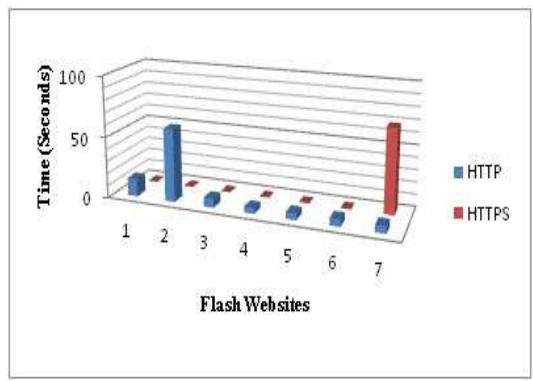}
\caption{Performance Analysis on Sony Ericson WT13i.}
\end{figure}

\begin{figure}[ht!]
\centering
\includegraphics[width=7cm, height=5cm]{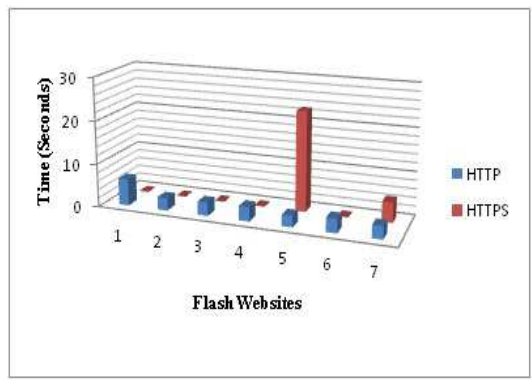}
\caption{Performance Analysis on LG Optimus Net.}
\end{figure}

\vskip 2mm

On SonyXPeria10 device C-Web pages takes time between 30 to 389.82 seconds, where as T-Web pages takes very less time between 1.06 to 2 seconds. Figure 8 depicts pictorial view of difference between response time of C and T Web pages. Similarly we have done the same analysis on Sony Ericson WT13i (Figure 9), here most of the C-Web pages are not displayed because it required Flash player, Java Script. Whereas, T-Web pages are gets displayed successfully between 4.77 to 59.56 seconds. Similar processes were carried out on LG-Optimus hand held device. Figure 10 depicts pictorial view of difference between response time of C and T Web pages. Two Web domains namely www.noleath.com and www.marvismint.com  get downloaded completely. It was also observed that very rarely Home Page (HP) of a few sites was getting downloaded and rests required Flash Player (FP) to display the contents. In our approach, T-Web pages were displayed between 2.61 to 6.21 seconds on LG Optimus Net and also very less response time on other hand held devices as shown in Figure 13. 

\vskip 2mm
After experimental analysis, we have performed the content coverage analysis based on human perception as against the system view. The algorithm adopts kappa[22] statistics to quantitatively measure the degree of the agreement how Web pages share the similarity terms. Kappa result ranges from 0 to 1. The higher the value of Kappa, the stronger the similarity. Kappa more than 0.7 typically indicates that similarity of two Web pages is strong. Kappa values greater than 0.9 are considered excellent. Boolean values are used to represent the content coverage (0 \begin{math} \rightarrow \end{math}100\% content loss, 0.5 \begin{math} \rightarrow \end{math} 50\% content loss, 1 \begin{math} \rightarrow \end{math} 0\% content loss). Figure 14 depicts the clear view of kappa value relation between human perception (Human View- HV) and system view (SV). 		

\begin{equation}
\tiny
k= \frac{Pr(a) - Pr(e)}{1 - Pr(e)}
\end{equation}

Where $Pr(a)$ is the relative observed agreement among raters, and Pr(e) is the hypothetical probability of chance agreement, using the observed data to calculate the probabilities of each observer randomly saying each category. If the raters are in complete agreement then $k$=1. If there is no agreement among the raters other than what would be expected by chance (as defined by [23][24] $Pr(e)$), $k$ = 0.
\begin{figure*}[!htbp]
\centering
\begin{center}
\includegraphics[width=12cm, height=8cm]{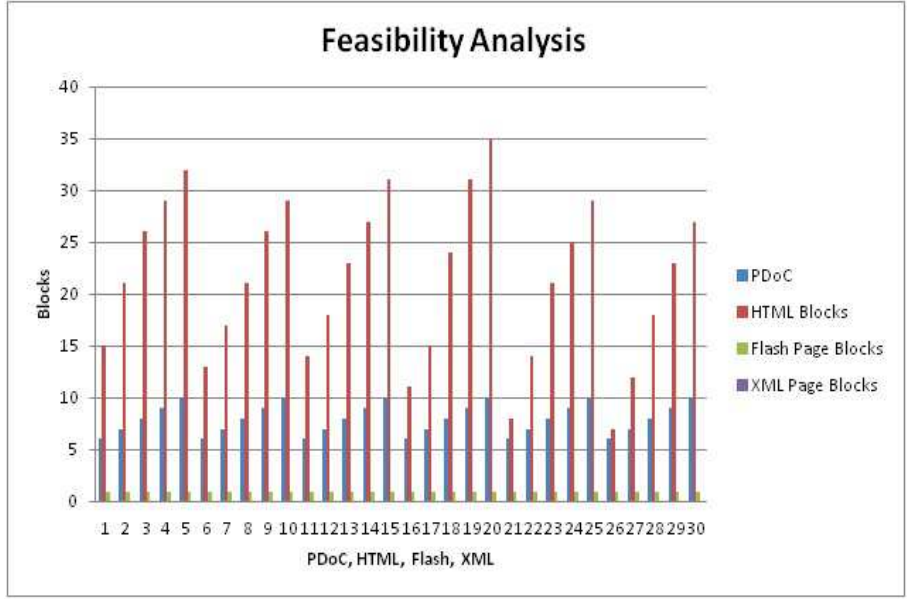}
\caption{Feasibility Analysis on Existing System.}
\end{center}
\end{figure*}

\begin{figure*}[ht!]
\centering
\begin{center}
\includegraphics[width=10cm, height=10cm]{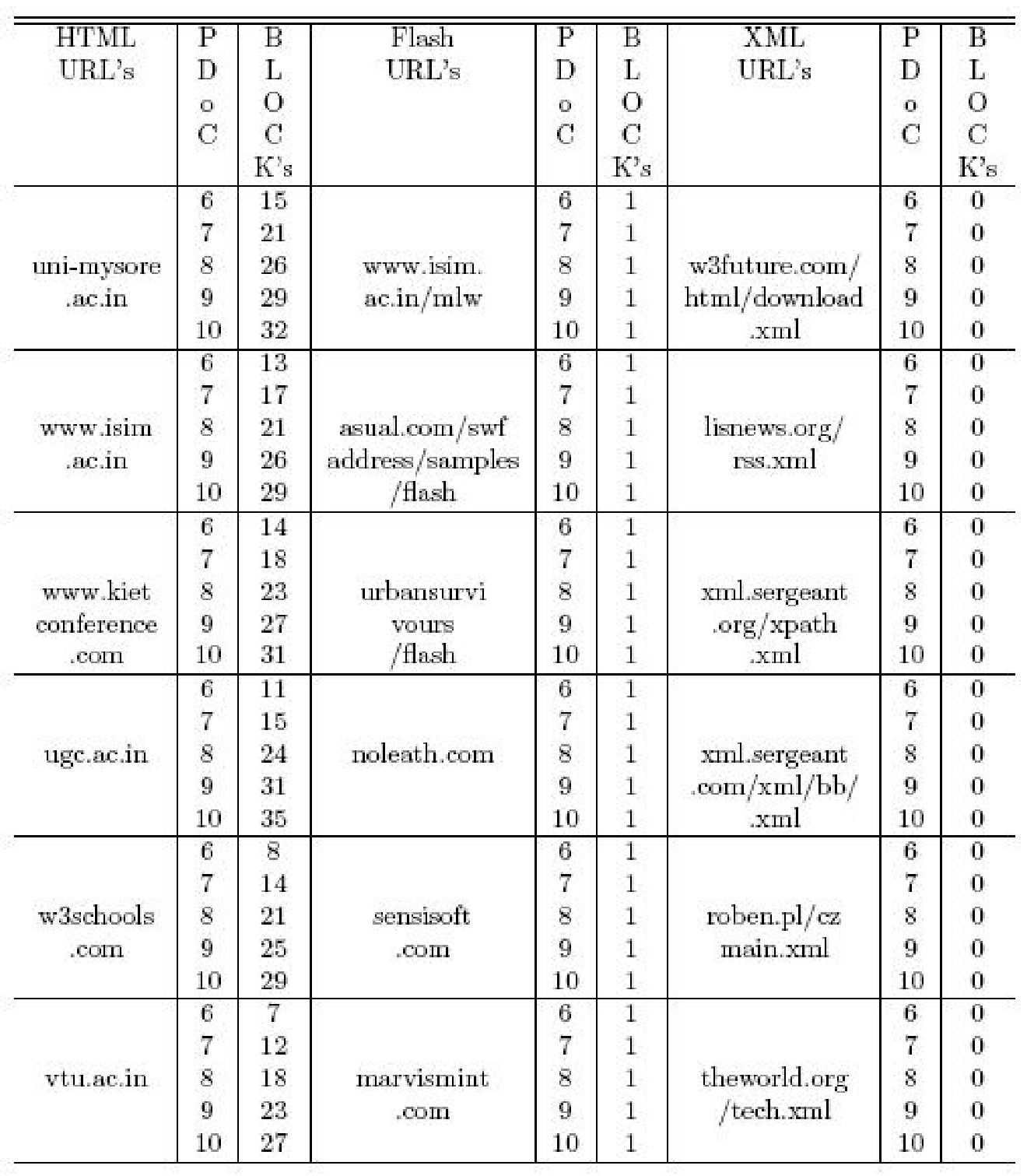}
\caption{Existing system Feasibility Analysis on our Data set.}
\end{center}
\end{figure*}

\begin{figure*}[ht!]
\centering
\begin{center}
\includegraphics[width=14cm, height=5.5cm]{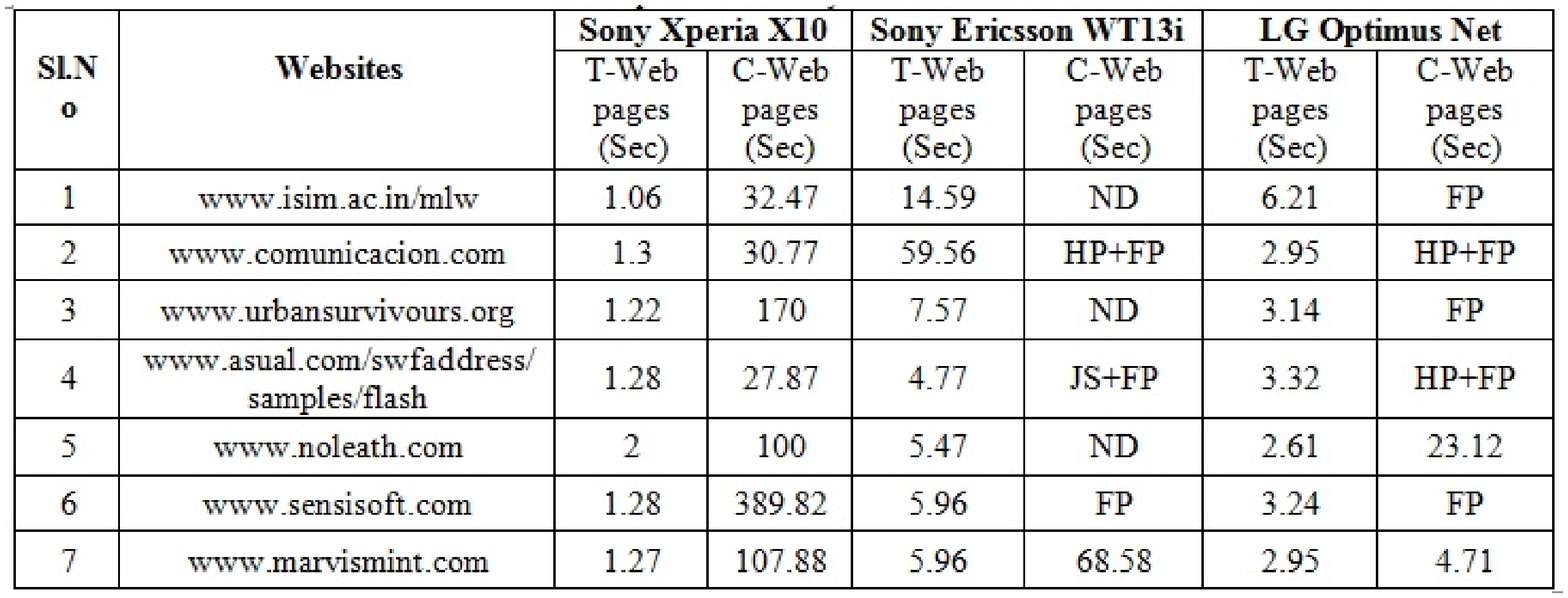}
\caption{Response Time Analysis on Various SSD’s.}
\end{center}
\end{figure*}

\begin{figure*}[ht!]
\centering
\begin{center}
\includegraphics[width=14cm, height=5.5cm]{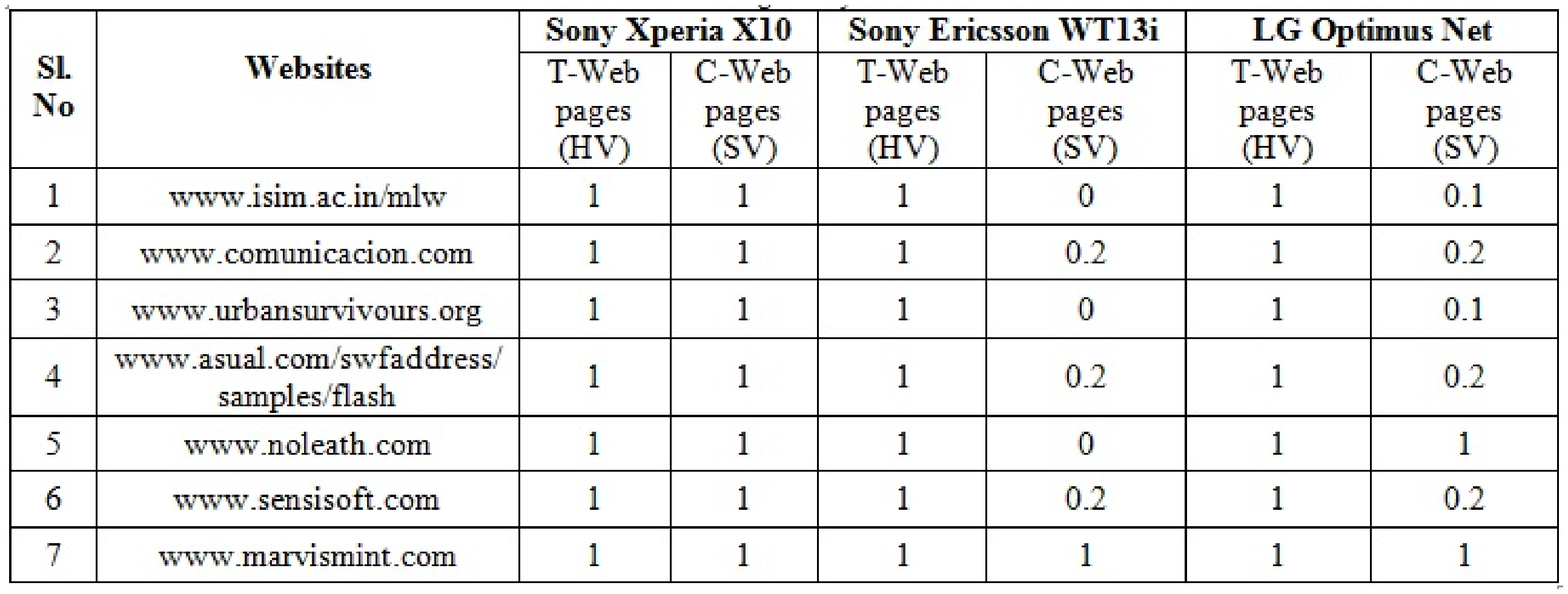}
\caption{Content Coverage Analysis on Various SSD’s.}
\end{center}
\end{figure*}

\section{CONCLUSIONS}
This paper presents an independent and comparative analysis of different existing approaches. From the experimental results of existing approaches, we claim that the performance of existing systems fail to provide result on our XML and Flash data sets. Subsequently, from the results obtained from our experiments we arrive at the following inferences: 1) Existing approaches fails to accommodate new popular technology web pages like XML, Flash $etc.$. 2) Existing approaches are purely designed for  HTML semantic structure orientation. Followed by this in this paper we have introduced the methodology to adopt Flash Web pages on SSD's and performed the analysis on various hand held devices. Here we achieved the better performance level based on response time and content coverage analysis. 

\small
\balance

\noindent{\includegraphics[width=1in,height=1.3in,clip,keepaspectratio]{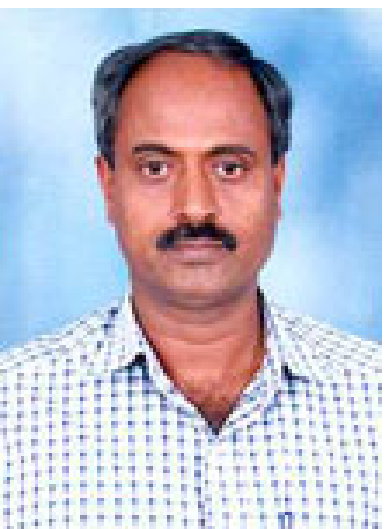}}
\begin{minipage}[b][1in][c]{1.8in}
{\centering{\bf{Suresha }}is a Professor, Dept. of Computer Science, University of Mysore, Mysore, India. He obtained his M.Sc from University of Mysore. M.phil from Devi Ahilya Viswa Vidyalaya, Indore, India. He received his M.Tech from Indian Institute of Technology Kanpur, India and  }\\\\
\end{minipage} Ph.D in Computer Science from Indian Institute of Science, Banglore, India. His area of research intrest is Web Technologies, E-Governance, DBMS and Distributed Systems.\\\\

\noindent{\includegraphics[width=1in,height=1.8in,clip,keepaspectratio]{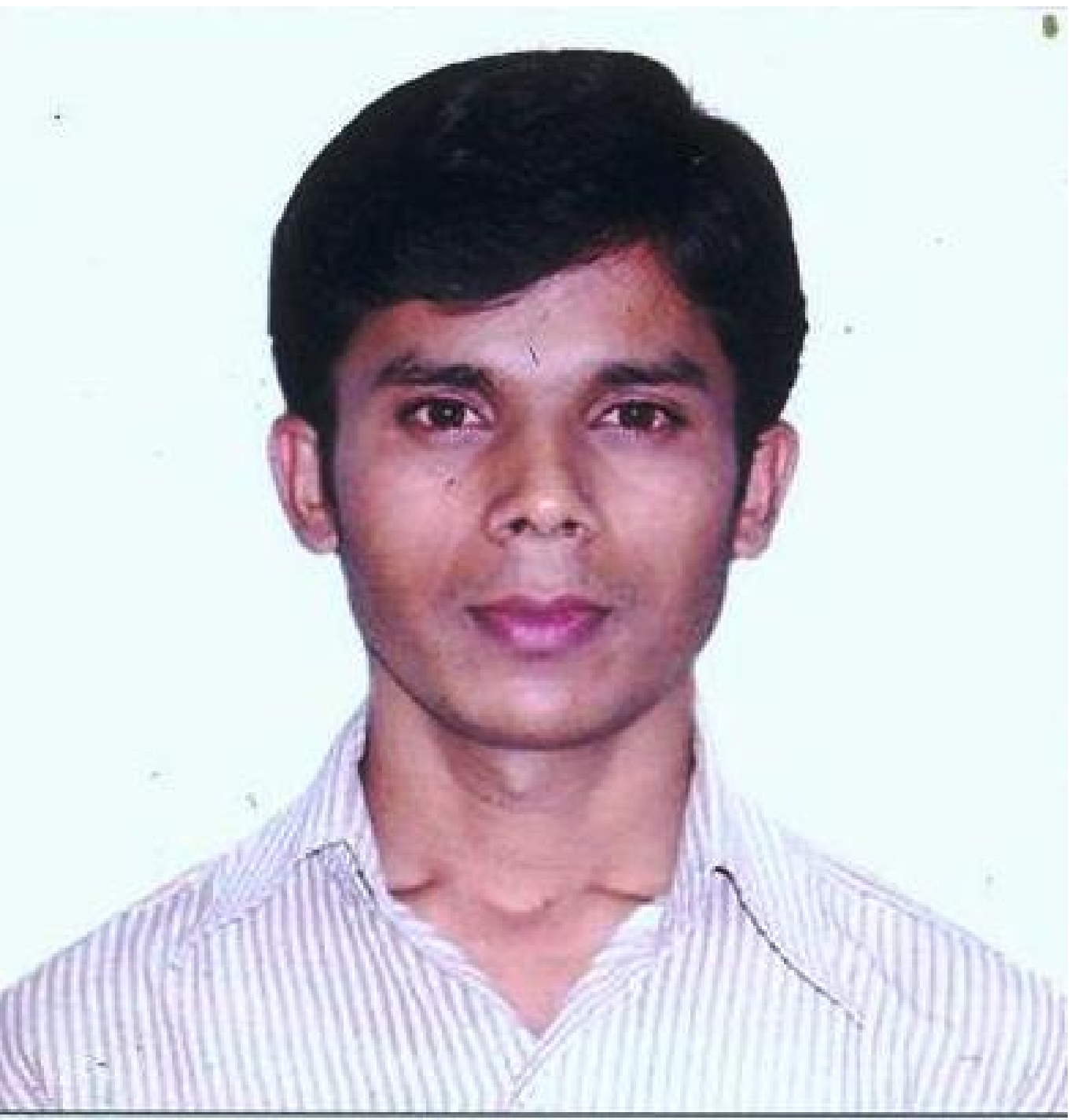}}
\begin{minipage}[b][1in][c]{1.7in}
{\centering{\bf {Krishna Murthy A}} is currently the Research Student working on Web Data Mining at University of Mysore, Mysore. He obtained his Bachelor of Science in Mathematics from Periyar University, Salem, Tamil Nadu}\\
\end{minipage} India. He received his Master degree in Computer Science from University of Mysore, Mysore, India.  \\\\
%---------------------------------------------------------------------------------------------------------------------
%\begin{biography}

\noindent{\includegraphics[width=1in,height=1.8in,clip,keepaspectratio]{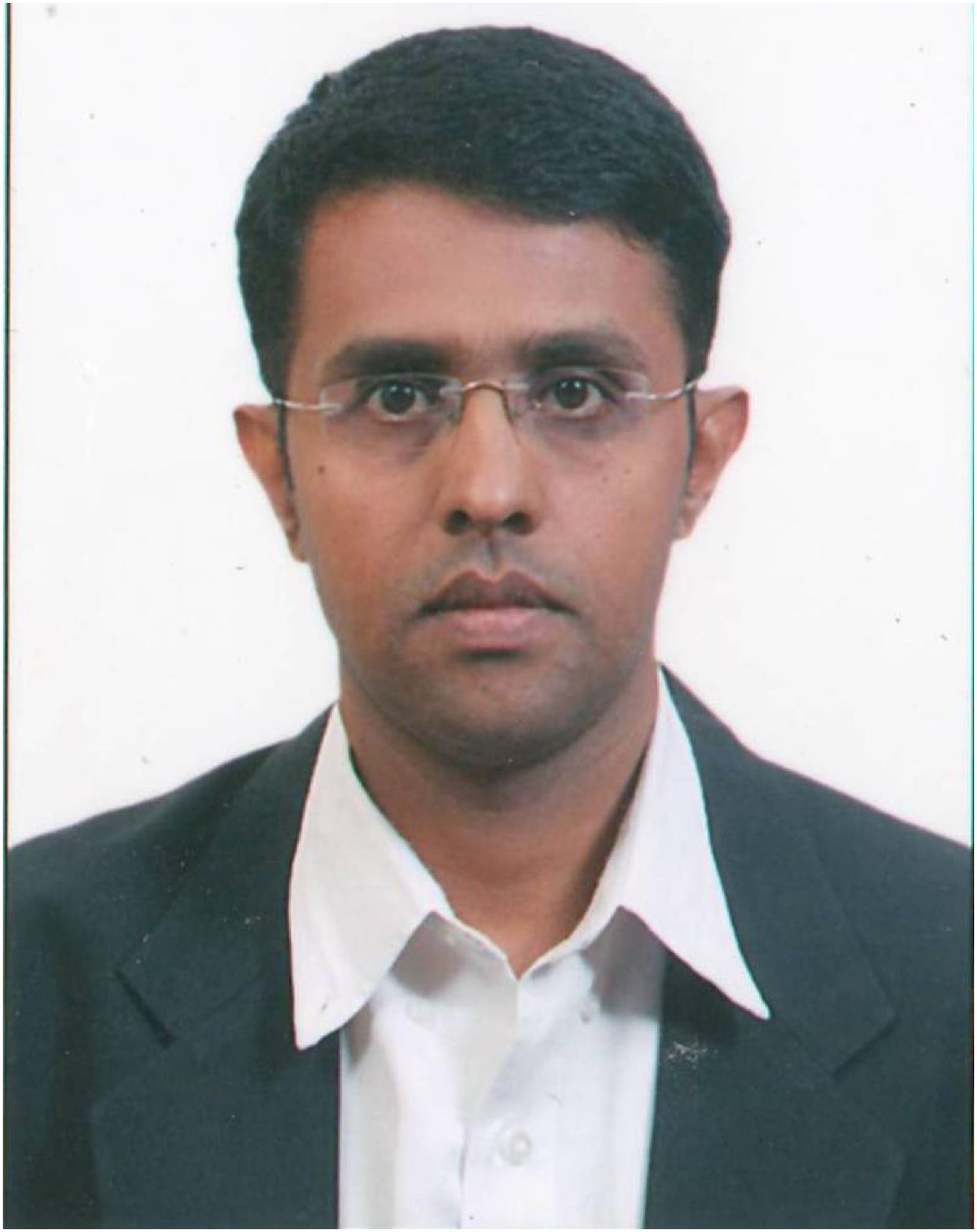}}
\begin{minipage}[b][1in][c]{1.7in}
{\centering{\bf {Anil Kumar K M}}  is currently working as Associate Professor, Dept of Computer Science,  Sri Jayachamarajendra College of Engineering, Mysore, India. Completed Master of Technology in Computer Network Engineering in 2006 from National Institute 
}\\\\
\end{minipage}  of Engineering, Mysore India. And he obtained his Doctoral of Philosophy in Computer Science in 2012 from University of Mysore, Mysore, India. His area of intrest is Web Tecchnologies, Analysis and Design of Algorithms, Network and Internet Technologies.  \\\\

\balance

\end{document}